\documentclass[lettersize,journal]{IEEEtran}
\usepackage{cite}
\usepackage{amsmath,amssymb,amsfonts}
\usepackage{graphicx}
\usepackage{textcomp}
\usepackage{algorithm}
\usepackage{subfigure}
\usepackage[caption=false,font=normalsize,labelfont=sf,textfont=sf]{subfig}
\usepackage{amsmath,amssymb} 
\usepackage{color}
\usepackage{multirow}
\usepackage{mathrsfs}
\usepackage{array}
\usepackage{threeparttable}
\usepackage{float}
\usepackage{cite}
\usepackage{algorithm}
\usepackage{algpseudocode}
\usepackage{eqparbox}
\usepackage{bm}
\usepackage{booktabs}
\usepackage{makecell}
\usepackage{diagbox}
\usepackage{textcomp}
\usepackage{stfloats}
\usepackage{url}
\usepackage{verbatim}
\usepackage{graphicx}
\def\BibTeX{{\rm B\kern-.05em{\sc i\kern-.025em b}\kern-.08em
		T\kern-.1667em\lower.7ex\hbox{E}\kern-.125emX}}
\usepackage{balance}

\begin{document}
	\title{Learning from Brain Topography: A Hierarchical Local–Global Graph-Transformer Network for EEG Emotion Recognition}
	\author{Yijin Zhou, 
		Fu Li$^*$,~\IEEEmembership{Member,~IEEE,}
		Yi Niu,
		Boxun Fu,
		Huaning Wang,
		Lijian Zhang
		\thanks{Yijin Zhou, Fu Li, Yi Niu, and Boxun Fu are with the Key Laboratory of Intelligent Perception and Image Understanding of Ministry of Education, the School of Artificial Intelligence, Xidian University, Xi’an, 710071, China.\it{($^*$Corresponding author: Fu Li (E-mail: fuli@xidian.edu.cn).)} \protect}
		\thanks{Huaning Wang is with the Department of Psychiatry at Xijing Hospital, Fourth Military Medical University, Xi’an, China.\protect}
		\thanks{Lijian Zhang is with the Beijing Institute of Mechanical Equipment, Beijing, 100000, China.\protect}		
		}
	\maketitle
	\begin{abstract}
		Understanding how local neurophysiological patterns interact with global brain dynamics is essential for decoding human emotions from EEG signals. However, existing deep learning approaches often overlook the intrinsic spatial organization of the brain and fail to simultaneously capture local topological relations and global temporal dependencies. To address these challenges, we propose Neuro-HGLN, a Neurologically-informed Hierarchical Graph–Transformer Learning Network that integrates biologically grounded priors with hierarchical representation learning. Neuro-HGLN first constructs a spatial Euclidean prior graph based on the physical distances between EEG electrodes, which serves as an anatomically grounded inductive bias. A learnable global dynamic graph is then introduced to model functional connectivity patterns across the entire brain, and a two-layer graph convolutional network is employed to extract global spatial representations. In parallel, to capture anatomically meaningful regional dependencies, Neuro-HGLN builds region-level local graphs and leverages a multi-head self-attention mechanism to enable fine-grained modeling of intra-regional spatial interactions. The local graphs corresponding to different brain regions are processed synchronously through local-constrained parallel GCN layers, producing region-specific representations. These local features are subsequently aggregated and fed into an iTransformer encoder, which captures cross-region dependencies and structured high-level feature interactions under the dimension-as-token formulation. Finally, a prediction head projects the fused representations into the emotion space. Extensive experiments demonstrate that Neuro-HGLN achieves state-of-the-art performance on multiple EEG emotion recognition benchmarks, while providing enhanced interpretability grounded in neurophysiological structure. Our results highlight the importance of unifying local topological learning with cross-region dependency modeling for robust and interpretable affective decoding from EEG signals.
	\end{abstract}

	
	\section{Introduction}
	\label{Introduction}
		Electroencephalography (EEG)–based emotion recognition has attracted increasing attention due to its potential applications in affective computing, human–computer interaction, and mental health assessment~\cite{wang2024research}. As a direct measurement of neural activity, EEG provides rich temporal information reflecting dynamic brain processes underlying emotional states~\cite{zheng2015investigating}. However, decoding emotions from EEG signals remains a challenging task, owing to the low signal-to-noise ratio, high inter-subject variability, and the complex spatial–temporal dependencies inherent in neural dynamics. 
	
		A key characteristic of EEG signals is their strong dependency on the spatial organization of the brain. The physical arrangement of electrodes reflects the underlying neuroanatomical structure, and the spatial proximity between electrodes often implies functional relevance~\cite{jang2021eeg}. Consequently, effectively modeling spatial relationships has been recognized as a critical factor for EEG emotion recognition. Graph-based learning methods have emerged as a promising paradigm, where EEG electrodes are treated as nodes and their interactions are modeled via graph structures. Recent studies have employed graph convolutional networks (GCNs) with predefined or learned adjacency matrices to capture spatial dependencies across channels~\cite{song2018eeg, song2021graph}. Despite their effectiveness, most existing approaches rely on either static graph priors or globally learned connectivity patterns, which may overlook fine-grained regional characteristics of brain activity.
		
		In addition to spatial dependencies, EEG signals exhibit complex temporal dynamics that span multiple time scales. Emotions are not instantaneous responses but evolve over time through interactions among distributed brain regions~\cite{li2025stereo}. While recurrent neural networks~\cite{zhang2018spatial} and temporal convolutional networks~\cite{zhao2026electrode} have been widely applied to model temporal dependencies, their ability to capture long-range interactions is often limited. Transformer-based models, with self-attention mechanisms, have recently demonstrated strong capability in modeling long-range dependencies in sequential data~\cite{vaswani2017attention}. Nevertheless, directly applying standard Transformers to EEG signals typically treats channels as independent tokens, ignoring the intrinsic spatial topology and hierarchical organization of the brain~\cite{arjun2021introducing}.
		
		To address these limitations, we propose Neuro-HGLN, a Neurologically-informed Hierarchical Local–Global Graph–Transformer Learning Network for EEG emotion recognition. Neuro-HGLN explicitly integrates biologically grounded spatial priors with hierarchical representation learning. Specifically, we construct a spatial Euclidean prior graph based on the physical distances between EEG electrodes, providing an anatomically meaningful inductive bias. On top of this prior, a learnable global dynamic graph is introduced to model functional connectivity across the entire brain, which is processed by graph convolutional layers to extract global spatial features. In parallel, Neuro-HGLN decomposes the electrode space into anatomically meaningful brain regions and constructs region-level local graphs. Through local-constrained parallel GCN layers and a multi-head self-attention mechanism, the model captures fine-grained intra-regional spatial dependencies. The resulting regional representations are then aggregated and fed into an iTransformer encoder~\cite{liu2023itransformer}, which captures cross-region dependencies and structured high-level feature interactions under the dimension-as-token formulation.
		
		Extensive experiments on multiple datasets demonstrate that Neuro-HGLN outperforms existing state-of-the-art methods in EEG emotion recognition. Beyond performance gains, the proposed framework offers improved interpretability by explicitly aligning the learning process with neurophysiological structures and spatial priors. These results highlight the importance of unifying local topological learning with cross-region dependency modeling for robust and interpretable affective decoding from EEG signals.
		
		The main contributions of this work are summarized as follows:
		
		$ \bullet $ We propose Neuro-HGLN, a neurologically informed hierarchical graph–transformer framework that jointly models local spatial topology and global cross-region feature dependencies for EEG emotion recognition.
		
		$ \bullet $ We introduce a spatial Euclidean prior graph together with a learnable global dynamic graph to capture anatomically grounded and functional connectivity patterns across EEG channels.
		
		$ \bullet $ We design a region-level local graph learning strategy with parallel GCN layers and attention mechanisms, enabling fine-grained modeling of intra-regional spatial dependencies.
		
		$ \bullet $ Extensive experimental results demonstrate that the proposed Neuro-HGLN framework achieves superior performance while offering enhanced interpretability grounded in neurophysiological structure.
		
		The rest of this paper is structured as follows. Section~\ref{Sec:Related works} summarizes existing approaches for EEG emotion recognition, with a focus on deep learning models, graph neural networks, and transformer-based architectures. Section~\ref{Sec: The proposed method} details the architecture and key components of the proposed Neuro-HGLN framework. Experimental settings and performance evaluations on multiple EEG emotion recognition benchmarks are presented in Section~\ref{Sec: Experiment}. Finally, Section~\ref{Sec: Conclusion} concludes the paper and outlines future research directions
		
	\section{Related Work}
	\label{Sec:Related works}
	
	\subsection{Deep Learning for EEG Emotion Recognition}
		Early approaches to EEG emotion recognition largely relied on traditional machine learning methods, such as Support Vector Machines (SVM) and Linear Discriminant Analysis (LDA), combined with manually extracted features like Differential Entropy (DE) and Power Spectral Density (PSD)~\cite{duan2013differential, chen2019feature, kilicc2022classification}. With the advent of deep learning, Convolutional Neural Networks (CNNs) and Recurrent Neural Networks (RNNs) have become dominant paradigms. For instance, zhao et al. utilized a dual-CNN framework—comprising a topology-aware 3D CNN and a computation-efficient 2D CNN—to systematically investigate the trade-off between electrode quantity and spatial configuration~\cite{zhao2026electrode}. Similarly, Zhang et al. proposed a Spatial–Temporal Recurrent Neural Network (STRNN) to unify feature learning from both domains. Distinct from standard approaches, they employed a multidirectional RNN layer to capture spatial co-occurrence by traversing spatial regions, followed by a bi-directional temporal RNN to model sequential dependencies. By further incorporating sparse projections to select salient regions, their framework demonstrates the capability of RNN-based architectures in simultaneously handling the complex spatial and temporal dynamics of EEG signals~\cite{zhang2018spatial}. 
		
		However, despite these advancements, inherent limitations remain. CNNs treat EEG signals as grid-like data (similar to images), which essentially ignores the non-Euclidean nature of brain topology. Similarly, while RNN-based methods (like STRNN) can integrate spatial information, they typically do so by flattening or traversing spatial dimensions sequentially. This implicitly models spatial dependencies rather than explicitly representing the complex, irregular connections between spatially distant electrodes, which is more naturally handled by graph structures.
		
	\subsection{Graph-based Networks for EEG Emotion Recognition}
		To explicitly model the inherently non-Euclidean spatial structure of EEG signals, Graph Neural Networks (GCNs) have been increasingly utilized in recent years. Specifically, EEG electrodes are represented as nodes, while their physical proximity or functional connectivity is encoded as edges, enabling structured spatial dependency modeling across channels.
		
		Existing GCN-based methods can be broadly categorized based on how the adjacency matrix is constructed. Prior studies have predominantly utilized static graph structures determined by the innate spatial topology of the EEG cap. In these approaches, the adjacency matrix is typically constructed based on the physical Euclidean distance between electrodes, where edge weights decay as the distance increases~\cite{jang2021eeg, zhong2020eeg}. Others proposed dynamic graph learning, where the adjacency matrix is learned from data to capture functional connectivity. For example, Song et al. introduced a Dynamical Graph Convolutional Neural Network (DGCNN) to overcome the limitations of static geometric graphs. Instead of relying on pre-defined physical distances, their method dynamically learns the adjacency matrix during the training process, allowing the network to capture the underlying functional connectivity between electrodes tailored for the specific emotion recognition task~\cite{song2018eeg}. Zhou et al. developed a progressive hierarchical graph convolutional network combining dynamic and static graphs to characterize both the intrinsic spatial proximity and dynamic functional connectivity of brain regions. Furthermore, inspired by the hierarchical nature of emotion, they introduced a dual-head module to progressively learn discriminative features from coarse-grained to fine-grained categories, thereby enhancing performance on complex emotion recognition tasks~\cite{zhou2023progressive}.
				
		Despite their efficacy, a critical limitation remains: most existing GCN architectures operate on a single-scale topology, performing global message passing across the entire electrode array indiscriminately. This formulation often neglects the intrinsic hierarchical modularity of the brain, failing to explicitly distinguish between dense intra-regional functional segregation and long-range inter-regional integration.
		
	\subsection{Transformers for EEG Emotion Recognition}
		
		While GCNs excel at spatial modeling, capturing long-range temporal dependencies in lengthy EEG signals remains challenging. Transformers, originally designed for natural language processing, have demonstrated superior capability in modeling long-range contexts via self-attention mechanisms. Several studies have adapted standard Transformers or Vision Transformers (ViT) to EEG signals. For instance, Xu et al. proposed the Attention-based Multiple Dimensions EEG Transformer (AMDET) to capture the global context of EEG signals, effectively utilizing a temporal attention block to identify critical time frames and model the long-term sequential dynamics that traditional spatial models often overlook~\cite{xu2023amdet}. Similarly, Arjun et al. applied Vision Transformers (ViT) to EEG analysis by converting raw signals into 2D time-frequency images via Continuous Wavelet Transform (CWT). By treating emotion recognition as an image classification task, their approach enables the model to leverage the global receptive field of ViT to extract robust emotional features from complex time-frequency representation~\cite{arjun2021introducing}.
		
		However, treating multi-channel EEG simply as a sequence of vectors can be suboptimal. To address this, the recently proposed iTransformer~\cite{liu2023itransformer} introduces a dimension-as-token strategy. By embedding each temporal series (channel) as a token, iTransformer effectively models the multivariate correlations and learns representations that are robust to temporal shifts, making it highly suitable for multi-channel biological signal processing.
		
		Nevertheless, applying Transformers directly often neglects the explicit spatial topology provided by biological priors. While Transformers can implicitly learn correlations, they lack the structural inductive bias that GCNs provide. Therefore, a hybrid architecture that unifies the structural biological priors of GCNs with the global dependency modeling of Transformers—specifically in a hierarchical local-global manner—remains an open and promising direction.
		
	\section{Methodology}
	\label{Sec: The proposed method}
	
\begin{figure*}[h]
			\centering {\includegraphics[width=1\linewidth]{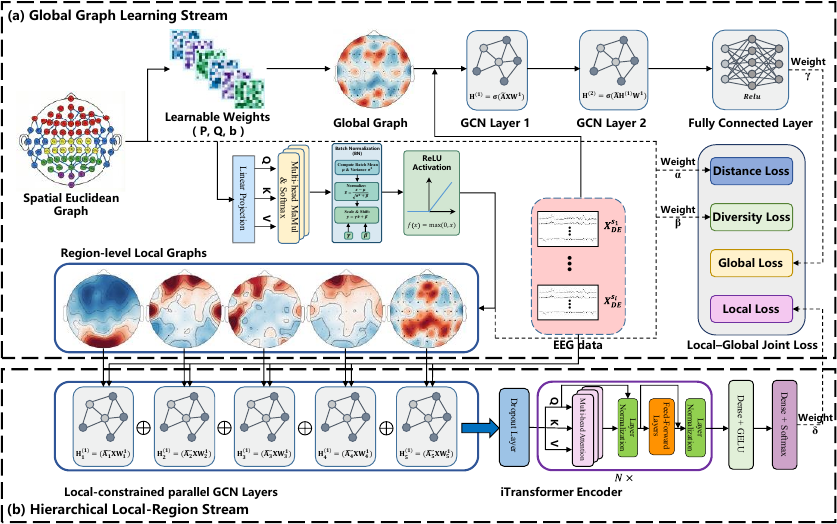}}
			\caption{\label{Neuro-HGLN framework} \textbf{The overall architecture of the proposed Neuro-HGLN framework.} The model orchestrates two synergistic streams to process EEG DE features: \textbf{(a) The Global Graph Learning Stream} captures holistic brain dynamics by integrating a static Spatial Euclidean Prior with learnable dynamic weights to construct a global graph, which is then processed by stacked GCN layers. \textbf{(b) The Hierarchical Local-Region Stream} models fine-grained dependencies by partitioning the brain into anatomical regions. It employs \textbf{Local-constrained Parallel GCNs} to extract intra-regional topology, followed by an \textbf{iTransformer Encoder} that captures high-level cross-region semantic dependencies using a dimension-as-token mechanism. The final emotion prediction is achieved by fusing the logits from both streams.
			}
		\end{figure*}

		To explicitly characterize the hierarchical topology of the brain for robust EEG-based emotion recognition, we propose the Neuro-HGLN framework, as illustrated in Fig.~\ref{Neuro-HGLN framework}. The core motivation of Neuro-HGLN is to emulate the neurological mechanism of \textit{functional segregation and integration}—where the brain processes information both within localized regions and through global communication across regions. Distinct from previous methods that rely on single-scale graph modeling or ignore spatial priors, Neuro-HGLN innovatively bridges the gap between biologically grounded spatial constraints and data-driven global dependency modeling. Through the synergistic fusion of local neurological priors and global semantic interactions, our method facilitates robust representation learning that strikes a balance between high accuracy and physiological interpretability
		
		As illustrated in Fig.~\ref{Neuro-HGLN framework}, the overall architecture comprises two synergistic streams designed to capture complementary neurological patterns:
		\begin{enumerate}
			\item \textbf{Global Graph Learning Stream:} This pathway captures holistic brain dynamics by constructing a global graph that integrates a static spatial Euclidean prior with learnable dynamic weights, enabling the model to respect anatomical constraints while adapting to functional connectivity.
			\item \textbf{Hierarchical Local-Region Stream:} This pathway models fine-grained regional characteristics. It utilizes parallel GCNs to extract intra-regional topological features and subsequently leverages an iTransformer encoder to model high-level cross-region dependencies under a dimension-as-token formulation.
		\end{enumerate}
		
	\subsection{Problem Formulation}
		
		Formally, let $\mathbf{X}_{raw} \in \mathbb{R}^{N \times T}$ denote the raw EEG signals, where $N$ represents the number of electrodes and $T$ is the number of time points. Following standard preprocessing protocols, we extract Differential Entropy (DE) features from the raw signals across five distinct frequency bands: $\delta$ (1--4 Hz), $\theta$ (4--8 Hz), $\alpha$ (8--14 Hz), $\beta$ (14--30 Hz), and $\gamma$ (30--50 Hz). 
		
		Consequently, the input to the network for a given subject $s_{i}$ is represented as the DE feature matrix $\mathbf{X}_{DE}^{s_i} = \{x_1, x_2, \dots, x_N\} \in \mathbb{R}^{N \times d}$, where $x_n \in \mathbb{R}^d$ denotes the feature vector of the $n$-th electrode and $d$ denotes the feature dimension. The goal of EEG emotion recognition is to learn a mapping function $\mathcal{F}: \mathbf{X} \to \mathbf{y}$, parameterized by $\Theta$, which projects the neural signals into the emotion space, where $\mathbf{y} \in \{1, \dots, C\}$ represents the predicted emotion category among $C$ classes.
	
	\subsection{Global Graph Learning Stream}
	\label{subsec:global_graph}
	
		To capture the holistic functional connectivity of the brain based on the anatomical structure, we construct a global graph $\mathcal{G}_g = (\mathcal{V}, \mathcal{E}_g, \mathbf{A}_g)$. While functional connectivity is dynamic, the physical proximity of electrodes imposes a fundamental constraint on signal propagation~\cite{salvador2005neurophysiological}. To incorporate this inductive bias, we explicitly construct a \textbf{Spatial Euclidean Prior Graph} based on standard 3D electrode positions.
		
		\subsubsection{Spatial Euclidean Prior Construction}
			Let $\mathcal{V} = \{v_1, v_2, \dots, v_N\}$ denote the set of $N$ EEG electrodes. We map each electrode $v_i$ to its standard 3D Cartesian coordinates $\mathbf{c}_i = (x_i, y_i, z_i) \in \mathbb{R}^3$ according to the International 10--10 System~\cite{NUWER20181103}. The physical distance matrix $\mathbf{D} \in \mathbb{R}^{N \times N}$ is computed using the Euclidean norm between electrode pairs:
			\begin{equation}
				\mathbf{D}_{ij} = \|\mathbf{c}_i - \mathbf{c}_j\|_2 = \sqrt{(x_i - x_j)^2 + (y_i - y_j)^2 + (z_i - z_j)^2}.
			\end{equation}
			
			To transform these physical distances into topological connection weights, we employ a Gaussian kernel function. This assumes that the strength of the neural interaction prior decays exponentially with the square of the physical distance. The static spatial adjacency matrix $\mathbf{A}_{prior} \in \mathbb{R}^{N \times N}$ is defined as:
			\begin{equation}
				\mathbf{A}_{prior, ij} = \exp\left(-\frac{\mathbf{D}_{ij}^2}{\tau}\right),
			\end{equation}
			where $\tau$ is a scaling factor (temperature parameter) that controls the sparsity and decay rate of the spatial connections. This formulation ensures that spatially adjacent regions have strong prior connections ($\mathbf{A}_{prior, ij} \to 1$ as $\mathbf{D}_{ij} \to 0$), while distant regions have weaker initial associations.
		
		\subsubsection{Dynamic Graph Fusion}
			However, physical distance alone is insufficient to capture the complex, non-linear functional coupling associated with varying emotional states. To address this, we introduce a learnable transformation mechanism to adaptively refine the rigid spatial priors. The final global adjacency matrix $\mathbf{A}_{g}$ is formulated as a fusion of the static prior and the learnable dynamic weights.
			
			Specifically, we utilize two learnable projection matrices $\mathbf{P}, \mathbf{Q} \in \mathbb{R}^{N \times N}$ and a bias matrix $\mathbf{b} \in \mathbb{R}^{N \times N}$ (corresponding to the parameters illustrated in Fig.~\ref{Neuro-HGLN framework}). The static spatial prior $\mathbf{A}_{prior}$ is projected into a dynamic functional space via the following transformation:
			\begin{equation}
				\mathbf{A}_{g} = \text{ReLU}\left( (\mathbf{P}\mathbf{A}_{prior} + \mathbf{b})\mathbf{Q} \right).
			\end{equation}
			In this formulation, $\mathbf{P}$ and $\mathbf{Q}$ facilitate the reshaping of the global connectivity patterns, while the bias $\mathbf{b}$ allows for the generation of intrinsic connections independent of physical distance. The ReLU activation function ensures the non-negativity of the edge weights, preserving the graph's structural validity.
			
			The resulting matrix $\mathbf{A}_{g}$ serves as the dynamic adjacency matrix for the subsequent graph convolution operations. The node representations at layer $l$, denoted as $\mathbf{H}^{(l)}$, are updated via standard GCN propagation:
			\begin{equation}
				\mathbf{H}^{(l+1)} = \text{ReLU}\left(\tilde{\mathbf{D}}^{-\frac{1}{2}} \tilde{\mathbf{A}}_g \tilde{\mathbf{D}}^{-\frac{1}{2}} \mathbf{H}^{(l)} \mathbf{W}^{(l)}\right),
			\end{equation}
			where $\tilde{\mathbf{A}}_g = \mathbf{A}_g + \mathbf{I}$ includes self-loops, and $\tilde{\mathbf{D}}$ is the degree matrix used for normalization.
			
			Following the stacked GCN layers, the final node representations, denoted as $\mathbf{H}^{(L)}$, contain high-level semantic features distributed across the topological structure. To map these features into the emotion space, we introduce a \textbf{Fully Connected (FC) Layer}. The output logits of the global stream, $\mathbf{y}_{global} \in \mathbb{R}^C$, are computed as:
			\begin{equation}
				\mathbf{y}_{global} = \text{Flatten}(\mathbf{H}^{(L)}) \mathbf{W}_{fc} + \mathbf{b}_{fc},
			\end{equation}
			where $\mathbf{W}_{fc}$ and $\mathbf{b}_{fc}$ denote the weight matrix and bias of the dense layer, respectively, and $C$ is the number of emotion classes. This output vector $\mathbf{y}_{global}$ effectively encapsulates the holistic spatial distribution of brain activity grounded in physical topology, serving as a distinct prediction branch that will be jointly optimized in the final loss function.
	
	\subsection{Hierarchical Local-Region Stream}
	\label{subsec:local_stream}
	
		While the Global Graph Stream captures holistic patterns, brain activity is fundamentally characterized by functional segregation—where distinct anatomical regions process information locally before integrating it globally~\cite{sporns2013network}. To emulate this mechanism, we propose the \textbf{Hierarchical Local-Region Stream} (as shown in Fig.~\ref{Neuro-HGLN framework} (b)), which orchestrates a two-stage learning process: intra-regional topological extraction followed by inter-regional semantic integration.
	
		\subsubsection{Intra-Regional Topology Learning via Parallel GCNs}
			First, based on standard neurophysiological partition schemes, we decompose the electrode set $\mathcal{V}$ into $K$ disjoint anatomical regions $\mathcal{R} = \{R_1, R_2, \dots, R_K\}$ (e.g., Frontal, Temporal, Occipital lobes). 
			
			Within each region $R_k$, we aim to capture fine-grained local topology. Instead of relying on fixed connections, we propose an \textbf{Attention-based Graph Proposal} mechanism that directly learns the optimal connectivity structure from the spatial prior to dynamically construct \textit{Region-level Local Graphs}.
			
			Similarly, we utilize the global spatial prior, $\mathbf{A}_{prior}$, as the initial topological structure for region $R_k$. To extract high-level structural features from this local prior, we apply a learnable transformation analogous to the global stream. Specifically, we project $\mathbf{A}_{prior}$ into a latent graph-feature space using region-specific parameters $\mathbf{P}_k, \mathbf{Q}_k$, and $\mathbf{b}_k$:
			
			\begin{equation}
				\mathbf{A}_{k} = \text{ReLU}\left( (\mathbf{P}_k \mathbf{A}_{prior} + \mathbf{b}_k)\mathbf{Q}_k \right),
			\end{equation}
			where $\mathbf{A}_{k} \in \mathbb{R}^{N \times N} $ represents the learned semantic embedding of the $ k $th intra-regional topology.
			
			Subsequently, to explore complex dependencies beyond the initial topology, we feed $\mathbf{A}_{k}$ into a Multi-Head Self-Attention (MSA) module. Uniquely, we utilize the graph embeddings themselves as queries, keys, and values ($\mathbf{Q}=\mathbf{K}=\mathbf{V}=\mathbf{A}_{k}$). The raw attention scores for the $h$-th head are computed as:
			\begin{equation}
				\mathbf{S}_h^k = \frac{(\mathbf{A}_{k}\mathbf{W}_Q^h)(\mathbf{A}_{k}\mathbf{W}_K^h)^T}{\sqrt{d_k}},
			\end{equation}
			where $\mathbf{W}_Q^h, \mathbf{W}_K^h$ are learnable weights for the attention head. 
			
			The final adjacency matrix $\mathbf{A}_{local}^k$ is derived not from the value projection, but directly from the attention weights. To ensure robust and non-negative connectivity, we apply Batch Normalization (BN) and ReLU activation to the attention scores, followed by aggregating across heads:
			\begin{equation}
				\mathbf{A}_{local}^k = \sum_{h=1}^{H} \text{ReLU}\left(\text{BN}\left(\text{Softmax}(\mathbf{S}_h^k)\right)\right).
			\end{equation}
			This generated $\mathbf{A}_{local}^k$ represents the refined, data-driven topology for region $R_k$, which is subsequently used to guide the local GCN aggregation.
			
			To process these local graphs efficiently while preserving their distinct topological characteristics, we design \textbf{Local-constrained Parallel GCN Layers}. As illustrated in Fig.~\ref{Neuro-HGLN framework}, $K$ independent GCN modules operate in parallel, preventing feature smoothing across unrelated regions at this early stage. This design choice effectively captures the immediate intra-regional structural information. For the $k$-th region, the feature update rule is defined as:
			\begin{equation}
				\mathbf{H}_{k} = \text{GELU}\left(\tilde{\mathbf{A}}_{local}^k \mathbf{X}_{DE} \mathbf{W}_{k}\right),
			\end{equation}
			where $\mathbf{X}_{DE}$ denotes the input DE features, $\mathbf{W}_{k}$ is the region-specific learnable weight matrix, and $\tilde{\mathbf{A}}_{local}^k$ represents the normalized local adjacency matrix derived from the previous attention step.

			Following this parallel local processing, the distinct regional representations are aggregated to reconstruct the holistic brain feature map. We concatenate the outputs from all $K$ regions:
			\begin{equation}
				\mathbf{H}_{local} = \text{Concat}(\mathbf{H}_1, \mathbf{H}_2, \dots, \mathbf{H}_K) \in \mathbb{R}^{N \times (K*d')},
			\end{equation}
			where $N$ is the total number of electrodes, and $d'$ denotes the transformed feature dimension output by the GCN projection. This reconstructed $\mathbf{H}_{local}$ integrates the fine-grained topological features from all sub-regions and serves as the input token sequence for the subsequent iTransformer encoder.		
			
		\subsubsection{Inter-Regional Integration via iTransformer}
		\label{subsubsec:itransformer}
			While parallel GCNs capture short-range dependencies, emotion regulation heavily relies on the long-range coordination between varying brain regions. Standard Transformers typically treat temporal steps as tokens, which is suboptimal for multivariate EEG data where spatial channel correlations are paramount.
			
			To address this, we adopt the \textbf{iTransformer} encoder structure with a ``Dimension-as-Token'' strategy. Specifically, we treat the feature vector of each electrode (channel) as an independent token. The aggregated local features $\mathbf{H}_{local}$ are fed into the encoder to explicitly model the multivariate correlations. Let $\mathbf{Z}^{(0)} = \mathbf{H}_{local}$. The encoder consists of $N$ stacked layers, each comprising Multi-Head Self-Attention (MSA) and Feed-Forward Networks (FFN). The update process for the $n$-th layer is formulated as:
			\begin{align}
				\mathbf{Z}' &= \text{LayerNorm}(\mathbf{Z}^{(n-1)} + \text{MSA}(\mathbf{Z}^{(n-1)})), \\
				\mathbf{Z}^{(n)} &= \text{LayerNorm}(\mathbf{Z}' + \text{FFN}(\mathbf{Z}')),
			\end{align}
			where $\mathbf{Z}^{(N)} \in \mathbb{R}^{N \times d_{model}}$ denotes the output of the final encoder layer. In this formulation, the self-attention mechanism acts as a dynamic, fully connected graph that integrates information across the entire brain topology, effectively capturing high-level inter-regional dependencies.
			
			To map these high-level inter-regional representations into the emotion space, we design a specific MLP Projection Head. First, the encoder output is projected into a hidden semantic space via a dense layer with GELU activation:
			\begin{equation}
				\mathbf{H}_{mlp} = \text{GELU}(\mathbf{Z}^{(N)} \mathbf{W}_{1} + \mathbf{b}_{1}),
			\end{equation}
			where $\mathbf{W}_{1} \in \mathbb{R}^{d_{model} \times d_{hidden}}$ and $\mathbf{b}_{1}$ are learnable parameters. Subsequently, the feature map is flattened to combine information from all channels and projected to the output dimension $C$ (number of emotion classes):
			\begin{equation}
				\mathbf{y}_{local} = \text{Flatten}(\mathbf{H}_{mlp}) \mathbf{W}_{2} + \mathbf{b}_{2}.
			\end{equation}
			Here, $\mathbf{y}_{local} \in \mathbb{R}^{C}$ represents the unnormalized local-view logits, which provide complementary information to the global stream and will be jointly optimized.

	\subsection{Local-Global Fusion and Joint Optimization}
	\label{subsec:optimization}
	
		To synthesize the complementary insights from the holistic global view and the fine-grained local view, we employ a decision-level fusion strategy. Furthermore, to ensure the learnable graph structures are both biologically plausible and functionally diverse, we introduce a joint optimization objective incorporating geometric constraints and diversity regularization.
		
		\subsubsection{Decision-Level Fusion}
			During the inference phase, the final emotion prediction is derived by aggregating the confidence scores from both streams. Let $\mathbf{y}_{global}$ and $\mathbf{y}_{local}$ denote the logits produced by the Global Graph Learning Stream (Sec.~\ref{subsec:global_graph}) and the Hierarchical Local-Region Stream (Sec.~\ref{subsec:local_stream}), respectively. The final fused prediction $\mathbf{y}_{final}$ is computed as the arithmetic mean of the logits:
			\begin{equation}
				\mathbf{y}_{final} = \frac{1}{2} \left( \mathbf{y}_{global} + \mathbf{y}_{local} \right).
			\end{equation}
			The predicted class label is then determined by $\hat{y} = \arg\max(\mathbf{y}_{final})$. This late-fusion strategy acts as an ensemble mechanism, stabilizing predictions by mitigating the variance inherent in individual streams.
		
		\subsubsection{Joint Loss Function}
			To train the framework end-to-end, we design a composite loss function $\mathcal{L}_{total}$ that orchestrates classification accuracy with structural regularization. The objective consists of three components:
		
			\paragraph{Deep Supervision Classification Loss}
				To prevent gradient vanishing and ensure that both streams learn discriminative features independently, we apply Deep Supervision. We compute the Cross-Entropy (CE) loss for both the global and local branches against the ground truth labels $y$:
				\begin{align}
					\mathcal{L}_{global} = \mathcal{L}_{CE}(\mathbf{y}_{global}, y), \\			
					\mathcal{L}_{local} =  \mathcal{L}_{CE}(\mathbf{y}_{local}, y),
				\end{align}
					
			\paragraph{Geometric Constraint (KL Divergence)}
				While the region-level local graphs are learnable, it is crucial that they do not deviate excessively from the physical anatomical structure, which acts as a strong inductive bias. We introduce a \textbf{Geometric Constraint Loss} using the Kullback-Leibler (KL) Divergence~\cite{naidji2026egdn}. This term encourages the learned local adjacency matrix $\mathbf{A}_{local}^k$ to approximate the distribution of the spatial prior $\mathbf{A}_{prior}$:
				\begin{equation}
					\mathcal{L}_{dist} = \sum_{k=1}^K \text{KL}\left( \mathbf{A}_{prior} \parallel \mathbf{A}_{local}^k \right).
				\end{equation}
				By minimizing $\mathcal{L}_{dist}$, we enforce a soft constraint that anchors the learned topology to the spatial reality of the brain, preventing the graph generation module from overfitting to noise.
			
			\paragraph{Functional Diversity Regularization}
				The brain is a highly complex system where different regions perform specialized functions (functional segregation). To prevent the parallel GCNs from collapsing into identical modes (i.e., learning redundant topological patterns), we propose a \textbf{Functional Diversity Loss}. We penalize the overlap between the adjacency matrices of distinct regions by minimizing their pairwise inner product:
				\begin{equation}
					\mathcal{L}_{div} = \sum_{i=1}^{K} \sum_{j=i+1}^{K} \left\| \mathbf{A}_{local}^i \odot \mathbf{A}_{local}^j \right\|_F^2,
				\end{equation}
				where $\odot$ denotes the element-wise product and $\|\cdot\|_F$ is the Frobenius norm. Minimizing this term promotes orthogonality among the learned graphs, ensuring that each GCN branch focuses on distinct, non-overlapping functional connectivity patterns.
			
			\paragraph{Total Objective}
				The final objective function is formulated as a weighted sum of four distinct components: the classification losses from both the global and local streams, and the two structural regularization terms. Formally, the total loss $\mathcal{L}_{total}$ is defined as:
				\begin{equation}
					\mathcal{L}_{total} = \alpha \mathcal{L}_{dist} + \beta \mathcal{L}_{div} + \gamma \mathcal{L}_{global} + \delta \mathcal{L}_{local},
				\end{equation}
				where $\mathcal{L}_{global}$ and $\mathcal{L}_{local}$ denote the Cross-Entropy losses for the global graph learning stream and the hierarchical local-region stream, respectively. The hyperparameters $\gamma$ and $\delta$ balance the contribution of the two prediction branches. Simultaneously, $\alpha$ and $\beta$ control the strength of the geometric constraint ($\mathcal{L}_{dist}$) and functional diversity regularization ($\mathcal{L}_{div}$), respectively. By minimizing $\mathcal{L}_{total}$, Neuro-HGLN jointly optimizes for accurate emotion recognition while ensuring the learned graph structures are both anatomically grounded and functionally distinct. The training procedure for Neuro-HGLN is outlined in Algorithm~\ref{alg:neuro_hgln}. Further implementation details are provided in Section~\ref{Experiment setting} \textit{2)}.
							
				\begin{algorithm}[h]
					\caption{Training Procedure of the Neuro-HGLN model.}
					\label{alg:neuro_hgln}
					\begin{algorithmic}[1]
						\Require EEG feature data batch $\mathbf{X}$, corresponding
						\Statex \ \ \ \ \ \ \ class labels $\mathbf{Y}$; Spatial Euclidean prior graph $\mathbf{A}_{prior}$.
						\Ensure Optimized model parameters $\Theta$.
						\State Initialize model parameters (GCNs, iTransformer, etc.);
						\For{each epoch $ e \in [1, E] $}
						\For{each batch $(\mathbf{X}_b, \mathbf{Y}_b)$ in training set}
						\State \textbf{Global Graph Learning Stream}
						\State Generate dynamic global graph $\mathbf{A}_{g}$ via learnable
						\Statex \ \ \ \ \ \ \ \ \ transformation of $\mathbf{A}_{prior}$;
						\State Extract global representations $\mathbf{H}^{(L)}$ via stacked
						\Statex \ \ \ \ \ \ \ \ \ GCNs and compute global logits $\mathbf{y}_{global}$;
						
						\State \textbf{Hierarchical Local-Region Stream}
						\State Organize node features based on the $K$ prior
						\Statex \ \ \ \ \ \ \ \ \ anatomical regions;
						\For{each region $ k \in [1, K] $}
						\State Generate local graph $\mathbf{A}_{local}^k$ via Attention-
						\Statex \ \ \ \ \ \ \ \ \ \ \ \ \ based Graph Proposal mechanism;
						\State Extract intra-regional features $\mathbf{H}_k$ via single-
						\Statex \ \ \ \ \ \ \ \ \ \ \ \ \ layer Local-constrained Parallel GCN;
						\EndFor
						\State Concatenate regional features $\mathbf{H}_{local} = \text{Concat}( $ 
						\Statex \ \ \ \ \ \ \ \ \  $ \mathbf{H}_1 \dots \mathbf{H}_K)$;
						\State Model cross-region dependencies via iTransformer
						\Statex \ \ \ \ \ \ \ \ \ encoder to obtain local logits $\mathbf{y}_{local}$;
						
						\State \textbf{Joint Optimization}
						\State Calculate classification losses $\mathcal{L}_{global}$ and $\mathcal{L}_{local}$
						\Statex \ \ \ \ \ \ \ \ \ via Cross-Entropy;
						\State Calculate geometric constraint losses $\mathcal{L}_{dist}$
						\Statex \ \ \ \ \ \ \ \ \ (KL Divergence) and functional diversity 
						\Statex \ \ \ \ \ \ \ \ \ regularization $\mathcal{L}_{div}$;
						\State Compute total loss $\mathcal{L}_{total} = \alpha \mathcal{L}_{dist} + \beta \mathcal{L}_{div} $
						\Statex \ \ \ \ \ \ \ \ \ $ + \gamma \mathcal{L}_{global} + \delta \mathcal{L}_{local} $;
						\State Update parameters $\Theta$ according to gradient descent
						\Statex \ \ \ \ \ \ \ \ \ algorithm;
						\EndFor
						\EndFor
					\end{algorithmic}
				\end{algorithm}
	
	\section{Experiments}
	\label{Sec: Experiment}
	
	\subsection{Experimental Settings}
	\label{Experiment setting}
	\subsubsection{Datasets}
	
		To comprehensively evaluate the effectiveness and robustness of Neuro-HGLN, we conduct experiments on four widely-recognized public benchmarks: SEED, SEED-IV, SEED-V, and MPED. These datasets cover a diverse range of emotion categories, varying from coarse-grained (3 classes) to fine-grained (7 classes) tasks.
	
		\textbf{SEED}~\cite{zheng2015investigating}: This is a fundamental benchmark for EEG emotion recognition developed by SJTU. The dataset involves 15 healthy subjects (7 males and 8 females) participating in three independent sessions. In each session, participants were exposed to 15 film clips designed to elicit three discrete emotional states: \textit{positive}, \textit{neutral}, and \textit{negative}. EEG signals were recorded using a 62-channel ESI NeuroScan system\footnote{https://compumedicsneuroscan.com/} at a sampling rate of 1000 Hz and subsequently downsampled to 200 Hz. Following the standard protocol, we utilize Differential Entropy (DE) features extracted from 1-second non-overlapping segments across five frequency bands ($\delta, \theta, \alpha, \beta, \gamma$) as the model input.
		
		\textbf{SEED-IV}~\cite{8283814}: As an evolution of SEED, this dataset introduces higher complexity by defining four emotional categories: \textit{happy}, \textit{sad}, \textit{fear}, and \textit{neutral}. Data collection followed a similar protocol with 15 subjects across three sessions, where each session comprised 24 trials (6 clips per emotion). The signals were acquired via the same 62-channel setup based on the international 10--20 system. Consistent with SEED, the raw EEG data were preprocessed and converted into DE feature sequences to capture dynamic emotional changes.
		
		\textbf{SEED-V}~\cite{liu2021comparing}: To further challenge the model's capability in distinguishing fine-grained emotions, we employ the SEED-V dataset. This dataset includes data from 20 subjects (10 males and 10 females) and targets five distinct emotions: \textit{happy}, \textit{sad}, \textit{fear}, \textit{disgust}, and \textit{neutral}. Each subject completed three sessions, with 15 movie clips presented per session. The preprocessing pipeline aligns with the SEED series, ensuring consistency in feature distribution and enabling a fair comparison of model performance across varying degrees of emotional granularity.
		
		\textbf{MPED}~\cite{song2019mped}: Distinct from the SEED family, MPED is a large-scale dataset covering seven detailed emotion categories: \textit{joy, funny, anger, fear, disgust, sadness}, and \textit{neutral}. It involves 30 participants watching 28 video stimuli. While the dataset provides multi-modal physiological signals, our study exclusively utilizes the 62-channel EEG data to verify the proposed method. A 5th-order Butterworth filter (1--100 Hz) was applied to remove artifacts. Unlike the experimental settings for SEED datasets, for MPED, we extracted \textbf{256-point Short-Time Fourier Transform (STFT) features} from the preprocessed signals using a 1-second non-overlapping sliding window. These features were computed across the standard five frequency bands to capture fine-grained time-frequency characteristics suitable for the 7-class classification task.
	
	\subsubsection{Implementation Details}
	\label{Implementation Details}
			
		The proposed Neuro-HGLN framework was implemented using the TensorFlow library and trained on a high-performance computing workstation equipped with an NVIDIA GeForce RTX 3090 GPU. To ensure data consistency across the SEED, SEED-IV, SEED-V, and MPED datasets, the number of input nodes is standardized to $N=62$, corresponding to the 62-channel electrode montage provided by the ESI NeuroScan System. The input features are decomposed into five standard frequency bands: $\delta$ (1--4 Hz), $\theta$ (4--8 Hz), $\alpha$ (8--14 Hz), $\beta$ (14--30 Hz), and $\gamma$ (30--50 Hz), resulting in an input feature dimension of $d=5$ for each node.
		
		Regarding the model architecture, the parameters for the Hierarchical Local-Region Stream are configured based on the complexity of the task and empirical tuning. Specifically, the iTransformer encoder consists of $N=5$ stacked layers. To capture complex multivariate dependencies, the model dimension is set to $d_{model} = 120$, and the number of attention heads is set to $H = 15$, ensuring that each head processes a subspace of dimension $d_k = 8$. The inner dimension of the Feed-Forward Network (FFN) is set to $d_{ff} = 512$. For the projection heads and classifiers, we utilize GELU activation to introduce non-linearity.
		
		During the training phase, the model is optimized using the Adam optimizer. Instead of a fixed learning rate, we employ a dynamic learning rate schedule with a warm-up strategy to stabilize the training process. The learning rate is calculated as:
		\begin{equation}
			lr = d_{model}^{-0.5} \cdot \min(step^{-0.5}, step \cdot warmup\_steps^{-1.5}),
		\end{equation}
		where $step$ refers to the current training step number, and $warmup\_steps$ is set to 4000. The batch size is set to 128 to optimize GPU memory utilization and gradient stability. The hyper-parameters governing the joint loss function are rigorously calibrated: the weights for the main classification tasks are set to $\gamma = 10$ and $\delta = 10$ to prioritize prediction accuracy, while the regularization terms are weighted as $\alpha = 1 \times 10^{-3}$ (for geometric constraints) and $\beta = 0.025$ (for functional diversity).

	\subsection{Experimental Results}
	\label{sec:experimental_results}
	
	\subsubsection{Subject-dependent Evaluation}
	\label{sec:subject_dependent}
	
		To demonstrate the efficacy of Neuro-HGLN in capturing personalized emotional patterns, we conducted extensive subject-dependent experiments on SEED, SEED-IV, SEED-V, and MPED. For the SEED series (SEED, SEED-IV, SEED-V), we followed a strict cross-session validation protocol. Specifically, the model is trained on data from one session (e.g., Session 1) and evaluated on a completely different session (e.g., Session 2 or Session 3). For the MPED dataset, which consists of a single session with 28 trials (4 trials for each of the 7 emotion categories), we allocated 21 trials (3 per emotion) for training and the remaining 7 trials (1 per emotion) for testing.
	
		\textbf{Baselines.} We compared our framework against a diverse set of state-of-the-art baselines, ranging from traditional machine learning (Linear SVM~\cite{suykens1999least}) to advanced deep learning approaches. These include dynamic graph methods (DGCNN~\cite{song2018eeg}, PGCN~\cite{zhou2023progressive}), and domain adaptation techniques (RGNN~\cite{9091308}, TANN~\cite{LI202192}). All baseline results are cited directly from literature or reproduced under identical settings.
		
		\textbf{Performance Analysis.} The quantitative comparisons are summarized in Table~\ref{Table: cross-session}. Neuro-HGLN consistently outperforms all competing methods across the four datasets. Specifically, on the \textbf{SEED series}, Neuro-HGLN exhibits exceptional stability against temporal signal drift. In the challenging \textit{Session 1 $\to$ Session 3} task (where the time gap is largest), our model achieves \textbf{94.64\%} on SEED and \textbf{84.74\%} on SEED-IV, surpassing the strong baseline PGCN by \textbf{3.67\%} and \textbf{4.04\%}, respectively. This significant improvement can be attributed to the proposed \textbf{Geometric Constraint (KL Divergence)}. By anchoring the learnable local graphs to the spatial Euclidean prior, Neuro-HGLN prevents the model from overfitting to the specific noise patterns of the source session, thereby ensuring that the learned topological features remain robust across different days.
		
		It is worth noting that Neuro-HGLN even surpasses domain adaptation methods (e.g., RGNN and TANN), which utilize unlabeled test data during training. For instance, on the SEED-IV \textit{Session 2 $\to$ Session 3} task, Neuro-HGLN achieves \textbf{85.45\%}, outperforming the domain adaptation method TANN (80.02\%) by a margin of over 5\%. This is a critical advantage in real-world applications where test data is often unavailable. The superiority of Neuro-HGLN here suggests that the \textbf{Functional Diversity Regularization} effectively forces the parallel GCNs to learn orthogonal, non-redundant functional patterns. This intrinsic regularization acts as a powerful generalization mechanism, reducing the domain gap without requiring explicit alignment strategies.
		 
		On the fine-grained \textbf{MPED dataset} (7 emotions), our model achieves the highest accuracy of \textbf{44.96\%}, outperforming PGCN (43.56\%) and TANN (39.82\%). While fine-grained classification is notoriously difficult due to the subtle boundaries between emotions (e.g., \textit{joy} vs. \textit{funny}), the \textbf{iTransformer} module in our hierarchical stream plays a vital role here. By treating spatial dimensions as tokens, it explicitly models the complex multivariate correlations required to disentangle these overlapping emotional states, proving the effectiveness of the "Dimension-as-Token" strategy in high-class-count scenarios.
	
		\begin{table*}[h]
			\centering
			\fontsize{8}{11}\selectfont
			\setlength{\tabcolsep}{2pt} 
			\caption{Subject-dependent classification performance (Accuracy / STD \%) on SEED, SEED-IV, SEED-V, and MPED.}
			{       
				\begin{tabular}{c|ccc|ccc|ccc|c}
					\hline
					\hline
					\multirow{3}{*}{Method}  &\multicolumn{10}{c}{\textbf{ACC / STD (\%)}} \cr
					\cline{2-11}
					&\multicolumn{3}{c}{Session 1 $\rightarrow$ Session 2}   &\multicolumn{3}{c}{Session 1 $\rightarrow$ Session 3}   &\multicolumn{3}{c}{Session 2 $\rightarrow$ Session 3} &\textendash \cr
					\cline{2-11}
					& \textbf{SEED} & \textbf{SEED-IV} & \textbf{SEED-V} & \textbf{SEED} & \textbf{SEED-IV} & \textbf{SEED-V} & \textbf{SEED} & \textbf{SEED-IV} & \textbf{SEED-V} & \textbf{MPED} \cr
					\hline  
					SVM~\cite{suykens1999least} & 51.62/11.80 & 59.81/16.61 & 41.56/15.36 & 50.23/11.99 & 59.50/17.07 & 41.87/15.18 & 44.65/11.90 & 67.77/10.94 & 43.40/16.76 & 32.39/09.53 \cr
					DGCNN~\cite{song2018eeg} & 86.96/09.24 & 67.47/13.31 & 69.93/11.16 & 87.42/08.59 & 66.38/12.02 & 63.50/09.92 & 90.96/07.95 & 74.19/14.07 & 71.54/14.89 & 32.37/06.08 \cr
					PGCN~\cite{zhou2023progressive} & 91.24/03.75 & 79.87/11.60 & 75.07/10.13 & 90.97/08.59 & 80.70/11.60 & 70.23/10.72 & 93.43/07.40 & 82.22/08.50 & 75.05/13.38 & 43.56/08.21 \cr
					RGNN$ ^{\dag} $~\cite{9091308} & 89.77/08.85 & 70.08/10.54 & 72.56/10.52 & 88.59/07.14 & 70.07/14.59 & 69.14/10.61 & 92.33/11.56 & 74.59/11.31 & 69.28/10.11 & 38.77/08.14 \cr
					TANN$ ^{\dag} $~\cite{LI202192} & 90.91/12.47 & 73.85/09.46 & 74.78/08.66 & 90.40/12.57 & 73.38/15.29 & 69.08/09.43 & 91.64/17.10 & 80.02/08.49 & 77.63/09.20 & 39.82/07.98 \cr
					\textbf{Neuro-HGLN} & \textbf{92.57/07.80} & \textbf{82.23/12.45} & \textbf{76.80/11.75} & \textbf{94.64/09.37} & \textbf{84.74/13.30} & \textbf{73.72/12.83} & \textbf{95.21/09.21} & \textbf{85.45/15.11} & \textbf{78.79/08.94} & \textbf{44.96/08.25}\cr
					\hline
					\hline
				\end{tabular}
			}
			\begin{tablenotes}
				\footnotesize
				\item $ ^{\dag} $ Methods that utilize unlabeled test data for training (Domain Adaptation).
			\end{tablenotes}
			\label{Table: cross-session}
		\end{table*}
	\subsubsection{Subject-independent Evaluation}
	\label{sec:subject_independent}
	
		Subject-independent emotion recognition poses a significantly greater challenge than subject-dependent tasks due to the inherent physiological variance and non-stationary distribution of EEG signals across different individuals. To evaluate the robustness of Neuro-HGLN in this rigorous scenario, we employed a Leave-One-Subject-Out (LOSO) cross-validation strategy on all four datasets.
		
		\textbf{Baselines.} Consistent with the subject-dependent experiments, we evaluate Neuro-HGLN against the same set of representative methods to ensure a fair comparison. These include the traditional machine learning method \textbf{SVM}~\cite{suykens1999least}, dynamic graph-based deep learning models \textbf{DGCNN}~\cite{song2018eeg} and \textbf{PGCN}~\cite{zhou2023progressive}, as well as state-of-the-art domain adaptation frameworks \textbf{RGNN}~\cite{9091308} and \textbf{TANN}~\cite{LI202192}.
		
		It is crucial to note that the domain adaptation methods (RGNN and TANN) utilize unlabeled data from the test subjects during training to bridge the domain gap. In contrast, our Neuro-HGLN operates in a strictly inductive manner, relying solely on source subject data without accessing any information from the target domain.
		
		\textbf{Performance Analysis.} As shown in Table~\ref{Table: Subject-independent}, Neuro-HGLN achieves state-of-the-art performance across all datasets, consistently outperforming both dynamic graph methods and domain adaptation frameworks. 
		
		On the SEED and SEED-V datasets, Neuro-HGLN demonstrates exceptional generalization capabilities, achieving \textbf{90.92\%} and \textbf{78.34\%} accuracy, respectively. Notably, on SEED-V, it surpasses the strongest baseline PGCN by a significant margin of nearly \textbf{7\%}. Different from the subject-dependent scenario where temporal drift is the main issue, the core challenge here is the massive physiological variance across individuals. While domain adaptation methods (e.g., RGNN) attempt to statistically align these divergent feature distributions, Neuro-HGLN adopts a distinct strategy by leveraging the \textbf{Geometric Constraint (KL Divergence)}. This constraint effectively utilizes the standard electrode configuration as a "Universal Anatomical Scaffold." By forcing the learned graphs to respect this common physical topology, the model is guided to learn generalizable functional patterns that are shared across the human population, rather than overfitting to the idiosyncratic functional connectivity of specific training subjects.
		
		On the MPED dataset, which represents the most difficult task (7 emotions), Neuro-HGLN remains the top-performing model with an accuracy of \textbf{28.56\%}, slightly outperforming the runner-up PGCN (28.39\%) and the domain adaptation method RGNN (27.95\%). Although the numerical margin appears narrow, this result is significant given the complexity of the 7-class problem. Existing methods like PGCN rely heavily on global graph structures, which tend to suffer from the "over-smoothing" problem—blurring the subtle boundaries between similar emotions (e.g., \textit{joy} vs. \textit{funny}). In contrast, our Hierarchical Local-Region Stream explicitly isolates and processes regional features before global integration. This design preserves the fine-grained local discrepancies that are critical for distinguishing high-entropy emotion categories, validating that a hierarchical "divide-and-conquer" architecture is more robust than a holistic one in fine-grained recognition tasks.
	
		\begin{table}[h]
			\centering
			\fontsize{8}{11}\selectfont
			\setlength{\tabcolsep}{3pt} 
			\caption{Subject-independent emotion recognition results (Accuracy / STD \%) on SEED, SEED-IV, SEED-V, and MPED.}
			{       
				\begin{tabular}{c|cccc}
					\hline
					\hline
					\multirow{2}{*}{Method} &\multicolumn{4}{c}{\textbf{ACC / STD (\%)}} \cr
					\cline{2-5}
					&\textbf{SEED} &\textbf{SEED-IV} &\textbf{SEED-V} &\textbf{MPED} \cr
					\cline{1-5}       
					SVM~\cite{suykens1999least} & 56.73/16.29 & 37.99/12.52 & 23.71/08.25 & 19.66/03.96 \cr
					DGCNN~\cite{song2018eeg} & 79.95/09.02 & 52.82/09.23 & 41.92/06.67 & 25.12/04.20 \cr
					PGCN~\cite{zhou2023progressive} & 83.28/11.09 & 76.96/07.19 & 71.40/09.43 & 28.39/05.02 \cr
					RGNN$ ^{\dag} $~\cite{9091308}& 85.30/06.72 & 73.84/08.02 & 66.28/16.71 & 27.95/03.05 \cr
					TANN$ ^{\dag} $~\cite{LI202192}  & 84.41/08.75 & 68.00/08.35 & 67.36/08.22 & 28.32/05.11 \cr
					\textbf{Neuro-HGLN}    & \textbf{90.92/09.49} & \textbf{79.30/09.11} & \textbf{78.34/10.68} &\textbf{28.56/04.67} \cr
					\hline
					\hline
				\end{tabular}
			}
			\begin{tablenotes}
				\footnotesize
				\item $ ^{\dag} $ Methods using unlabeled test data (Domain Adaptation).
			\end{tablenotes}
			\label{Table: Subject-independent}
		\end{table}
	
	\section{Discussion}
	\label{sec:discussion}
	
		In this section, we conduct ablation studies to verify the contribution of each component in Neuro-HGLN. Furthermore, we visualize the learned topological structures and feature distributions to provide an intuitive interpretation of the model's inner workings.
	
	\subsection{Ablation Study}
	\label{subsec:ablation}
	
		To validate the effectiveness of the proposed modules, we conducted ablation experiments on the SEED-IV and MPED datasets by removing key components from the full model. We designed four variants:
		\begin{itemize}
			\item \textbf{w/o Global:} Removing the Global Graph Learning Stream, relying only on the local-region stream.
			\item \textbf{w/o Local:} Removing the Hierarchical Local-Region Stream (including parallel GCNs and iTransformer).
			\item \textbf{w/o Geo:} Removing the Geometric Constraint ($\mathcal{L}_{dist}$) to test the impact of the spatial prior.
			\item \textbf{w/o Div:} Removing the Functional Diversity Regularization ($\mathcal{L}_{div}$) to assess the redundancy of learned graphs.
		\end{itemize}
		
		The results are summarized in Table~\ref{Table: ablation}. We observe that:
		1) \textbf{Impact of Dual-Stream Architecture:} Removing either the Global (w/o Global) or Local (w/o Local) stream leads to a performance degradation. Notably, \textit{w/o Local} causes a sharper drop on the fine-grained MPED dataset compared to SEED-IV. This confirms our hypothesis that the hierarchical local stream is crucial for capturing subtle, fine-grained emotional patterns, while the global stream provides the necessary holistic context.
		2) \textbf{Impact of Regularization:} The removal of the Geometric Constraint (\textit{w/o Geo}) results in unstable performance, particularly in cross-session tasks, suggesting that the spatial prior acts as an essential regularizer against overfitting. Furthermore, \textit{w/o Div} leads to a noticeable decline, indicating that without diversity enforcement, parallel GCNs tend to collapse into learning redundant features, limiting the model's representation capacity.
		
		\begin{table}[h]
			\centering
			\fontsize{8}{11}\selectfont
			\setlength{\tabcolsep}{3pt} 
			\caption{Ablation study results (Accuracy \%) on SEED-IV and MPED.}
			\label{Table: ablation}
			\begin{tabular}{l|cc}
				\hline
				\textbf{Model Variant} & \textbf{SEED-IV (Avg.)} & \textbf{MPED} \\
				\hline
				\textbf{Full Model (Neuro-HGLN)} & \textbf{85.14} & \textbf{44.96} \\
				\hline
				~~ w/o Global Stream & 82.10 & 41.50 \\
				~~ w/o Local Stream  & 80.35 & 38.20 \\
				~~ w/o Geo. Constraint ($\mathcal{L}_{dist}$) & 83.90 & 43.10 \\
				~~ w/o Funct. Diversity ($\mathcal{L}_{div}$) & 84.15 & 42.80 \\
				\hline
			\end{tabular}
		\end{table}
		
	\subsection{Visualization of Learned Topology}
	\label{subsec:vis_topology}
		
\begin{figure}[htp]
			\centering 
			\subfigure[Global Graph]{\includegraphics[width=0.32\linewidth]{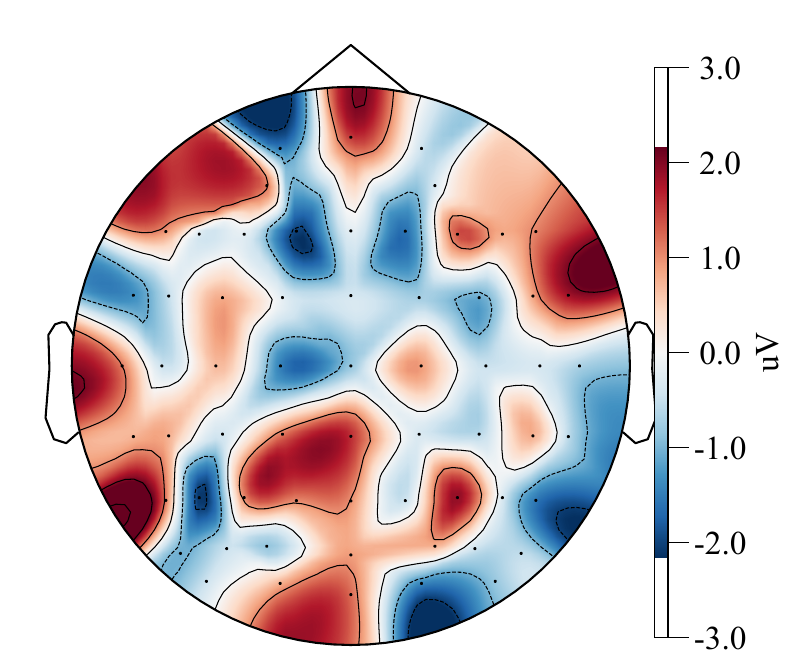}}
			\subfigure[Prefrontal]{\includegraphics[width=0.32\linewidth]{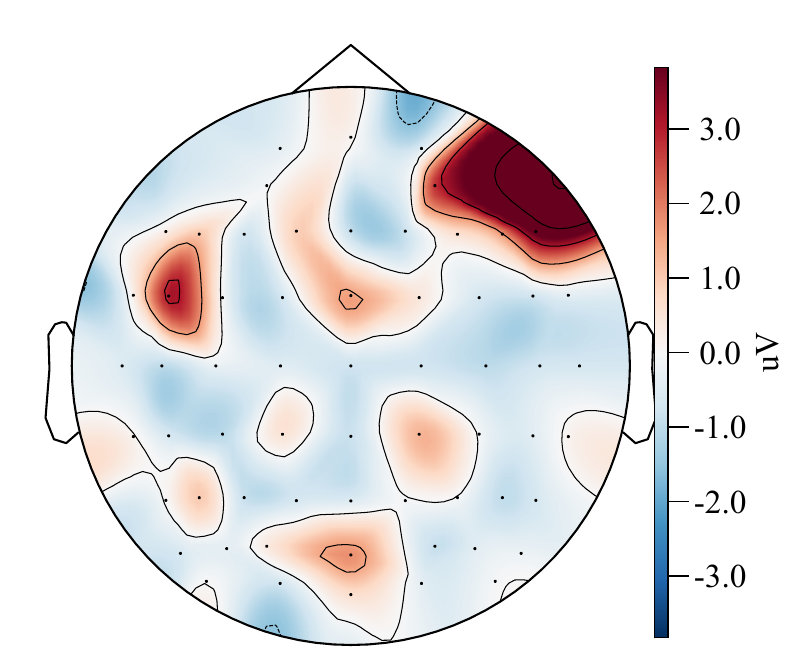}}
			\subfigure[L-Temporal]{\includegraphics[width=0.32\linewidth]{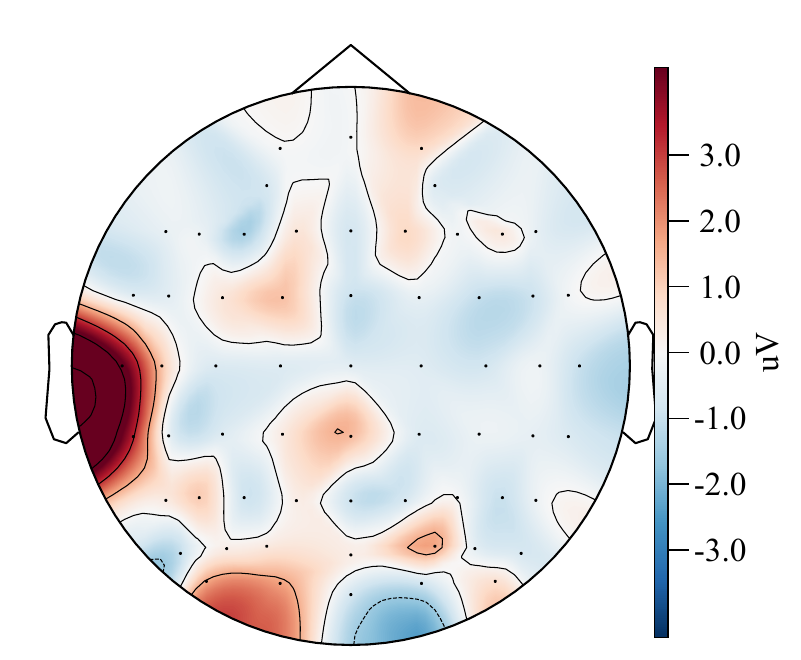}}
			\subfigure[R-Temporal]{\includegraphics[width=0.32\linewidth]{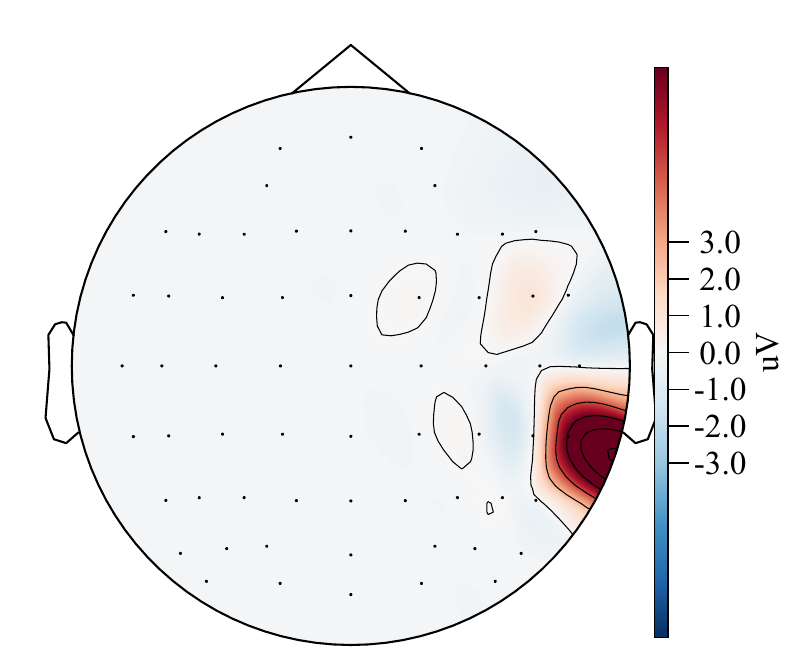}}
			\subfigure[Parietal]{\includegraphics[width=0.32\linewidth]{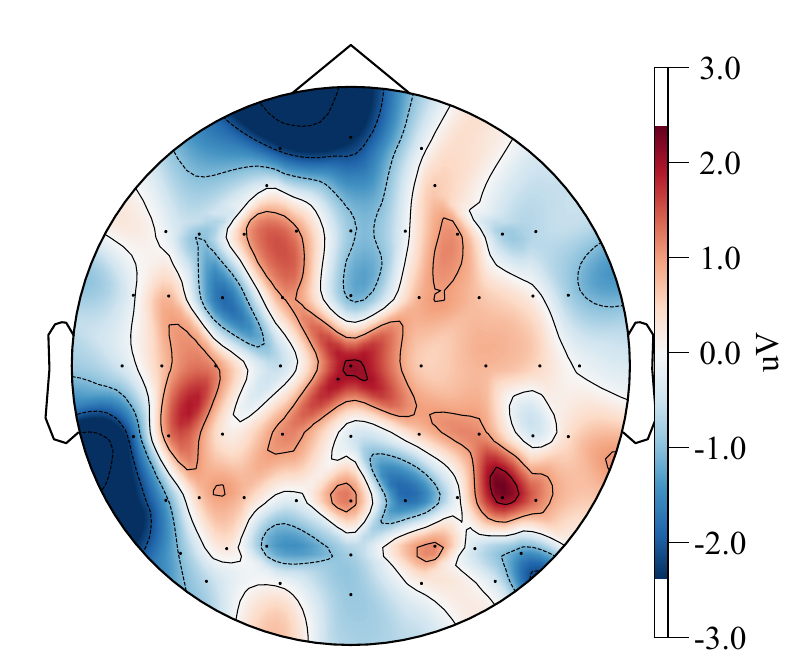}}
			\subfigure[Occipital]{\includegraphics[width=0.32\linewidth]{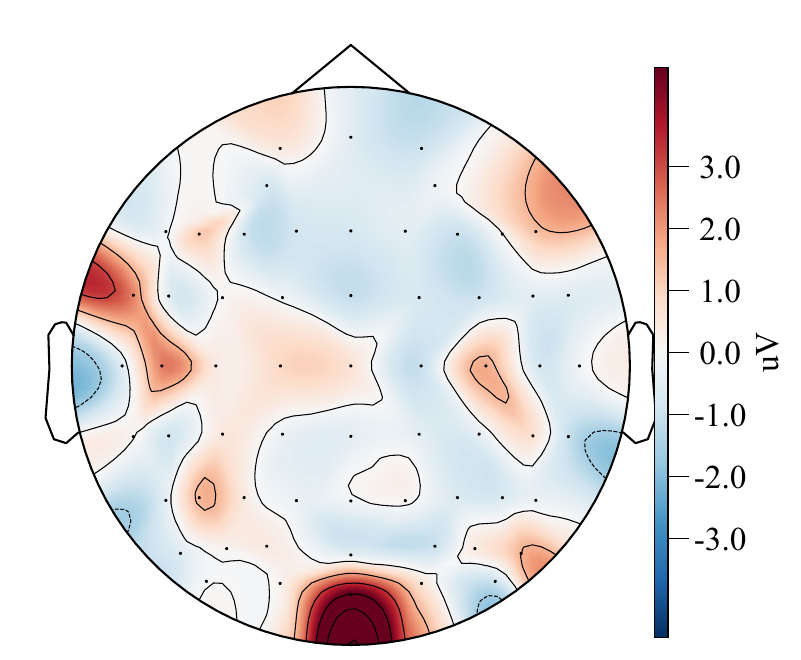}}
			\caption{\textbf{Visualization of learned topological structures.} (a) The Global Graph captures holistic, long-range dependencies. (b-f) The Local Graphs exhibit distinct, region-specific activation patterns (e.g., Prefrontal focus vs. Occipital focus), verifying that the Functional Diversity Regularization effectively prevents feature homogenization.}
			\label{fig:vis_topology}
		\end{figure}

		A core contribution of Neuro-HGLN is the ability to learn biologically plausible and functionally diverse connections. To verify this, we visualize the topological activation patterns learned by the Global Stream and the distinct Local Streams in Fig.~\ref{fig:vis_topology}.
		
		As illustrated in Fig.~\ref{fig:vis_topology}~(a), the \textbf{Global Graph} exhibits a widespread, complex activation pattern that spans across bilateral hemispheres. This distributed topology indicates that the global stream captures long-range dependencies and holistic brain dynamics, consistent with the lateralization theory of emotion processing~\cite{davidson1992anterior}.
		
		In sharp contrast, the \textbf{Local Graphs} (Fig.~\ref{fig:vis_topology}~(b-f)) demonstrate highly localized and region-specific patterns. Specifically, the \textbf{Prefrontal} map focuses intensely on the anterior regions , aligning with its role in emotion regulation and higher-order cognition~\cite{ochsner2005cognitive}. The \textbf{Temporal} maps (Left and Right) show distinct focal activations on the lateral sides, while the \textbf{Occipital} map exhibits a concentrated response in the posterior visual cortex~\cite{olson2007enigmatic, bradley2001emotion}.
		
		This distinctiveness validates the effectiveness of the \textbf{Functional Diversity Regularization} ($\mathcal{L}_{div}$). Without this regularization, these regional sub-graphs would tend to converge towards a homogenized pattern similar to the global graph. The clear structural separation observed here confirms that Neuro-HGLN successfully disentangles the functional roles of different anatomical regions, allowing the model to integrate diverse neurological perspectives for robust emotion recognition.
	
	\subsection{Feature Distribution Visualization}
	\label{subsec:tsne}
		To intuitively assess the discriminative power of the learned representations, we employ t-SNE~\cite{van2008visualizing} to project the high-dimensional features extracted by PGCN and Neuro-HGLN on the MPED dataset (7 emotion categories) into a 2D space.
		
\begin{figure}[h]
			\centering 
			\subfigure[PGCN]{\includegraphics[width=0.48\linewidth]{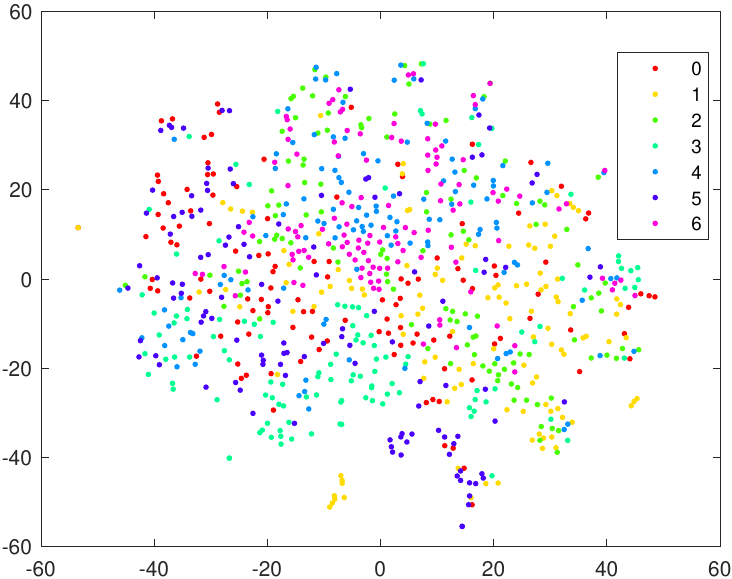}}
			\subfigure[Neuro-HGLN]{\includegraphics[width=0.48\linewidth]{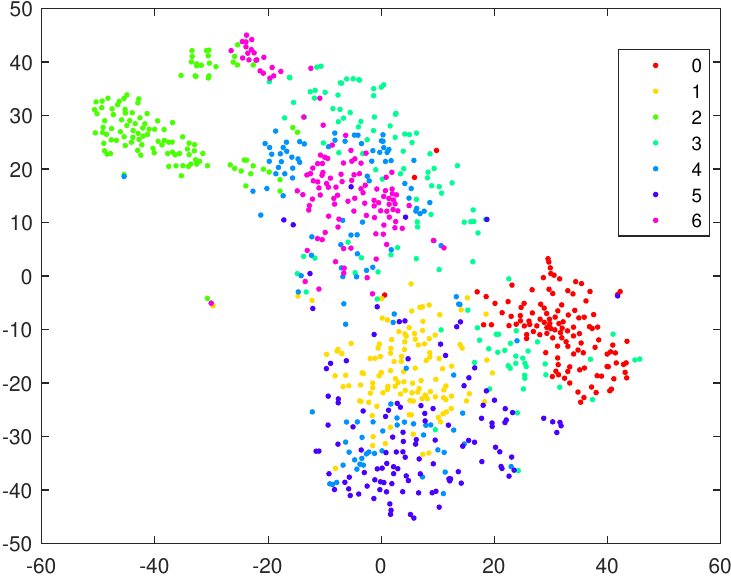}}
			\caption{\label{fig:tsne}\textbf{t-SNE visualization of feature distributions on the MPED dataset.} The digit labels 0--6 correspond to \textit{neutrality, joy, funny, anger, fear, disgust}, and \textit{sadness}, respectively. (a) PGCN shows a scattered distribution with severe class overlap, indicating poor separability. (b) Neuro-HGLN forms distinct, compact clusters with clear inter-class margins, demonstrating its superior ability to disentangle fine-grained emotional features.}
		\end{figure}

		As visualized in Fig.~\ref{fig:tsne}, the difference in feature quality is striking. PGCN (Fig.~\ref{fig:tsne}~(a)) suffers from severe feature entanglement, where samples from different emotion categories (represented by distinct colors) are mixed into a chaotic, unstructured cloud. This confirms that relying solely on global graph learning is insufficient to distinguish the subtle boundaries of fine-grained emotions.
			
		In contrast, Neuro-HGLN (Fig.~\ref{fig:tsne}~(b)) produces highly discriminative manifolds. The learned features exhibit significantly tighter intra-class clustering and clearer inter-class margins. Notably, several distinct clusters (e.g., the \textit{funny} and \textit{neutrality} groups) are completely isolated from the rest, indicating that our model has successfully captured the unique topological signatures of these emotions. Even for the confusing emotion pairs that overlap in the center, Neuro-HGLN maintains a much more organized structure than PGCN. This demonstrates that the integration of the Hierarchical Local-Region Stream effectively disentangles complex emotional states, projecting them into a separable latent space.
	
	\section{Conclusion}
	\label{Sec: Conclusion}
	
		In this paper, we propose \textbf{Neuro-HGLN}, a novel biologically grounded framework for robust EEG emotion recognition. By bridging the gap between global functional connectivity and local anatomical specialization, our method effectively addresses the challenges of physiological instability and cross-subject variability. Specifically, we introduce a \textbf{Hierarchical Local-Region Stream} that employs parallel GCNs to extract fine-grained, region-specific features, regularized by a Functional Diversity loss to prevent redundancy. Simultaneously, the \textbf{Global Graph Learning Stream} captures holistic brain dynamics, anchored by a \textbf{Geometric Constraint} that utilizes the spatial Euclidean prior as a "Universal Anatomical Scaffold." This constraint significantly enhances the model's generalization capability against temporal drift and individual differences without requiring domain adaptation techniques. Extensive experiments on four benchmark datasets (SEED, SEED-IV, SEED-V, and MPED) demonstrate that Neuro-HGLN achieves state-of-the-art performance in both subject-dependent and subject-independent tasks, particularly excelling in fine-grained emotion recognition scenarios.
		
		\textbf{Future Work.} While Neuro-HGLN shows promising results, several avenues remain for exploration. First, we plan to integrate \textbf{Dynamic Causal Modeling (DCM)} to infer directed information flow between brain regions, moving beyond undirected correlations to understand the causality of emotional processing. Second, we aim to extend our framework to \textbf{asynchronous multimodal learning}, incorporating eye-movement or physiological signals to further enhance robustness in real-world affective computing applications.
		
	\section{Acknowledgments}
		This work was supported in part by the STI 2030—Major Projects 2022ZD0208500, in part by the National Natural Science Foundation of China (NSFC) under Grants 62506282, 62376203, 62206210, 62371356, and 82330043, in part by the Key Research and Development Program of Shaanxi under Grant 2023-ZDLSF-07, in part by the Xidian University Specially Funded Project for Interdisciplinary Exploration under Grant TZJH2024021, in part by the Scientific Research Program Funded by the Shannxi Provincial Education Department under Grant 20JY022, in part by the Fundamental Research Funds of the Central Universities of China under Grants ZYTS23065, QTZX2107, RW200141, and 2242023k30022, and in part by the Project funded by China Postdoctoral Science Foundation under Grants 2024M762547, 2021M692504, and 2021TQ0259.
	
	\clearpage
	\bibliographystyle{IEEEtran}
	\bibliography{refbib.bib}
	
	\clearpage
	
	\begin{IEEEbiography}[{\includegraphics[width=1in,height=1.25in,clip,keepaspectratio]{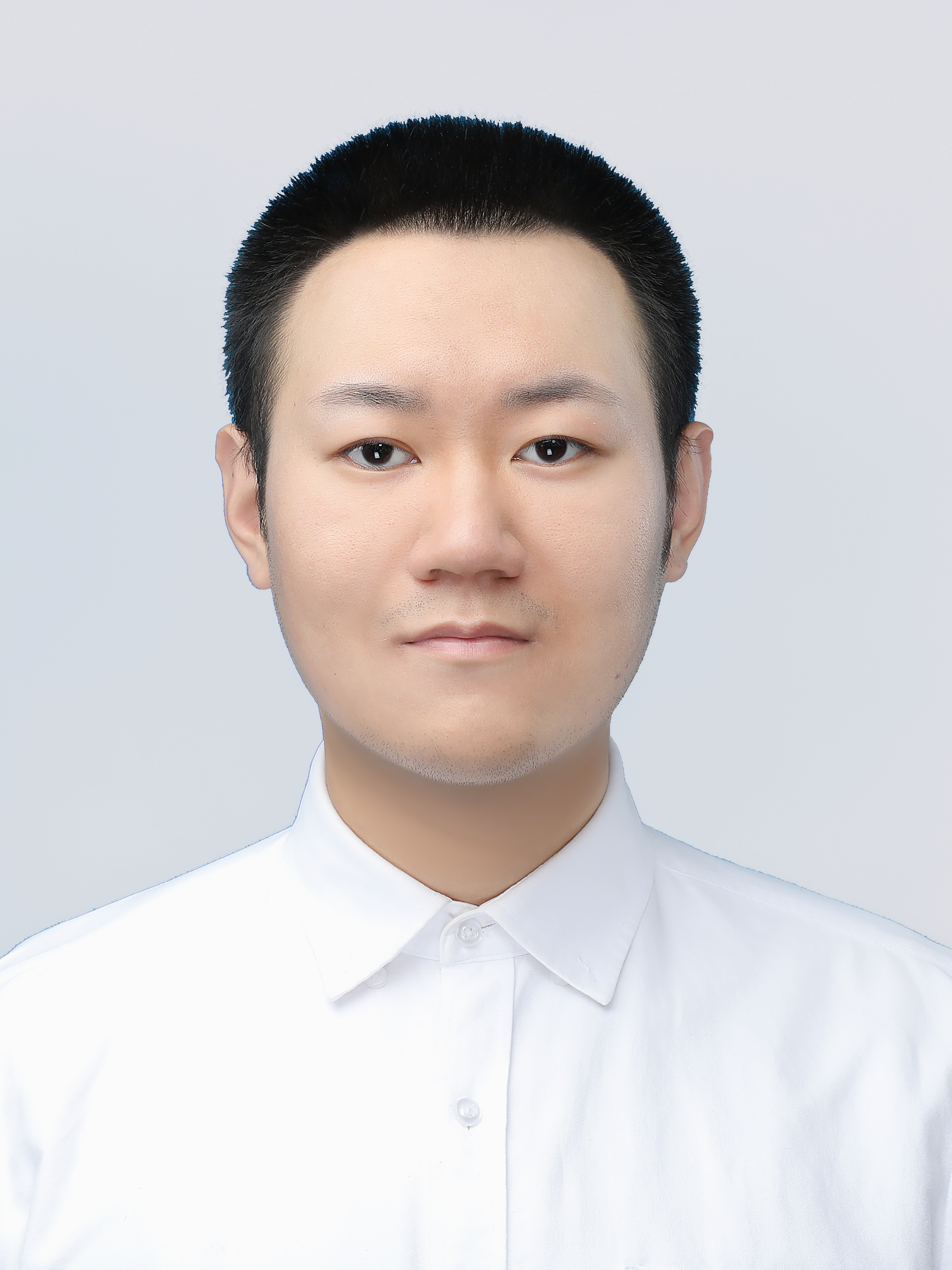}}]{Yijin Zhou} received the B.S. degree in Electronic and Information Engineering from Xidian University, Xi’an, China, in 2020. He is currently a Ph.D. candidate in the School of Artificial Intelligence, Xidian University, Xi’an, China. His research interests include affective computing, brain computer interface, computer vision, and machine learning, etc.
	\end{IEEEbiography}
	
	\begin{IEEEbiography}[{\includegraphics[width=1in,height=1.25in,clip,keepaspectratio]{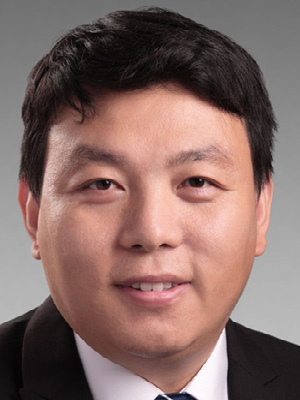}}]{Fu Li} is a Professor in the School of Artificial Intelligence at Xidian University, Xi’an, China. He is the head of Xidian-Xilinx Embedded Digital Integrated System Joint Laboratory. He received his B.S. degree in Electronic Engineering from Xidian University in 2004, and Ph.D. degree in Electrical \& Electronic Engineering from the Xidian University in 2010. He has published more than 30 papers international and national journals, and international conferences. His research interests are brain-computer interface, deep learning, small target detection, 3D imaging, embedded deep learning, image and video compression processing, VLSI circuit design, target tracking, neural network acceleration and Implementation of intelligent signal processing algorithms (DSP \& FPGA).
	\end{IEEEbiography}
	
	\begin{IEEEbiography}[{\includegraphics[width=1in,height=1.25in,clip,keepaspectratio]{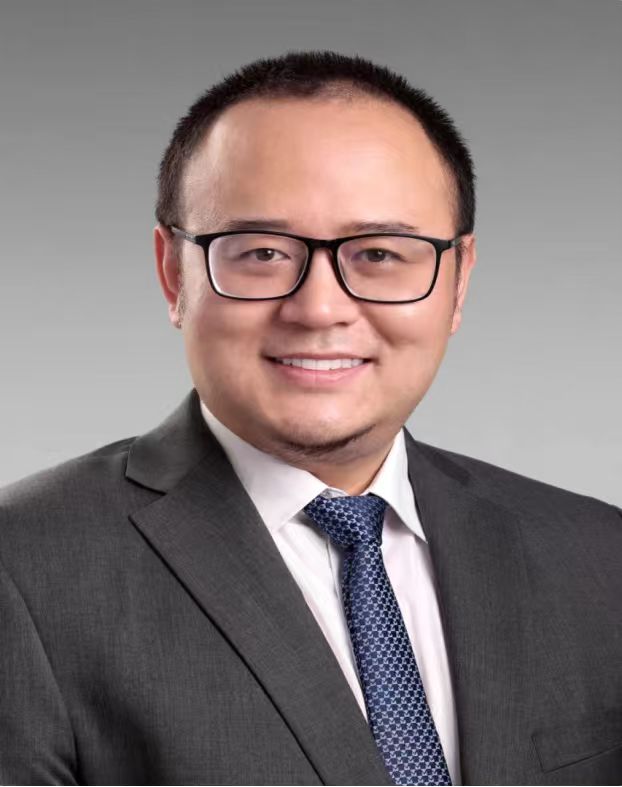}}]{Yi Niu} received the B.S. and Ph.D. degrees in electronic engineering from Xidian University, Xi'an,China, in 2005 and 2012, respectively. From 2009 to 2012, he was a Visiting Student with the Departmentof Electrical and Computer Engineering, McMasterUniversity, Hamilton, ON, Canada, where he was a Postdoctoral Fellow, from 2012 to 2013. He is currently a Professor with the School of Artificial Intelligence, Xidian University.
	\end{IEEEbiography}
	
	\begin{IEEEbiography}[{\includegraphics[width=1in,height=1.25in,clip,keepaspectratio]{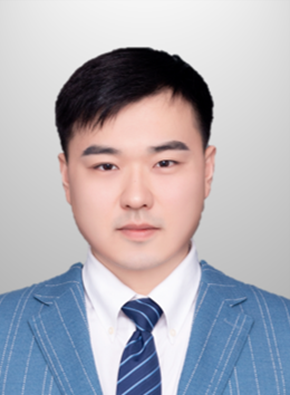}}]{Boxun Fu} received the B.S. degree in electronic engineering from Xidian University in 2016, and Ph.D. degree in Computer Science \& Technology from Xidian University in 2023. His research interests include brain computer interface, computer vision, and machine learning.
	\end{IEEEbiography}
	
	\begin{IEEEbiography}[{\includegraphics[width=1in,height=1.25in,clip,keepaspectratio]{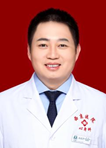}}]{Huaning Wang} received the Ph.D. degree from Fourth Military Medical University in 2009. His research interests focus on Physical therapy for mental disorders and neuromodulation.
	\end{IEEEbiography}
	
	\begin{IEEEbiography}[{\includegraphics[width=1in,height=1.25in,clip,keepaspectratio]{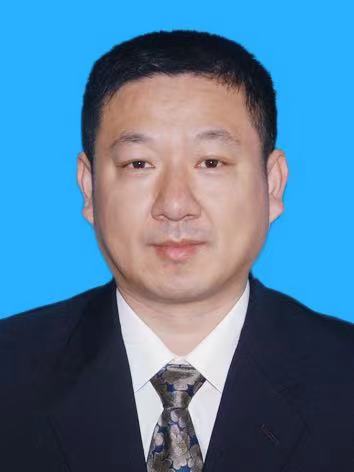}}]{Lijian Zhang} received the B.S. degree from North China University of technology in 1993, and the M.S. degree from Beijing Institute of technology in 2005. His research interests focus on Biomechanical, signal acquisition, biomedical signal processing, especially the study of brain-computer interface.
	\end{IEEEbiography}
	
\end{document}